\documentstyle[12pt]{article} 
\begin{document}
\addtolength{\baselineskip}{1mm}
\begin{flushright}
\end{flushright}
\begin{center}
{\Large\bf Comment II on J.M. Salazar and L. Brenig's contribution:\\
``Computer Simulations  and Kinetic Theory
of an Inelastic Granular Gas''}
\\[2ex]
{\large\it M.H. Ernst  \\
Institute for Theoretical Physics, P.O.Box 80.006 \\
3509 TA Utrecht, The Netherlands; E-mail:ernst@phys.uu.nl}
\end{center}

\vspace{1cm}

\begin{abstract}
\noindent
In two recent articles Salazar and Brenig question the validity of 
kinetic theory for granular gases and fluids, on the based of a supposedly
exact hierarchy  of coupled equations for the velocity moments, which 
the authors derive from the BBGKY-hierarchy. Their derivation contains several 
errors, which are exposed. 
Moreover, they support their findings with results from direct monte carlo 
simulation (DSMC) of the Boltzmann equation, which supposedly
show that Haff's homogeneous cooling law only holds for times shorter than 
one mean free time. However, their DSMC results have no physical significance 
as they are carried out at a density, twice close packing. 
\end{abstract}

\noindent
The recent comment in cond-mat/9911276 by Y. Elskens is completely in line 
with my criticism, voiced at the workshop on Granular Gases, that
took place in Bad Honnef on March 8 - 12, 1999 (see Ref. \cite{badhonnef}). 
It prompted the publication of the  notes, below made at that conference, 
which raise a number of additional objections.

Salazar and Brenig in Ref. \cite{brenig} claim to have obtained the following 
analytic and simulation  results for  undriven
(freely cooling) granular gases of Inelastic Hard Spheres (IHS): 

\begin{itemize}
\item[(i)] A Direct Simulation Monte Carlo  (DSMC) solution of the Boltzmann
equation, demonstrating that
the validity of Haff's law \cite{haff} is restricted to times less than one 
mean
free time $t_0$. This law states that the decay of the total energy from an 
initially homogeneous equilibrium state obeys $E(t) = 
E(0)/(1+t/t_\epsilon )^2$, where $t_\epsilon = 2d t_0/(1- 
\alpha^2)$  \cite{haff}
in a $d$-dimensional IHS fluid, and $\alpha$ is the coefficient of normal
restitution.

\item[(ii)] An exact analytic relation for the time dependence of certain
velocity moments, on the basis of which they question the validity of the
kinetic theory of granular gases.
\end{itemize}
In the present comment I show that their analytic relation does not hold, 
and that  their DSMC results refer to unphysical densities of twice close 
packing, and consequently have no physical relevance.

Concerning point (i): many authors have shown by means of molecular 
dynamics \cite{us,mcnam}, and of DSMC \cite{DSMC} that Haff's law is 
in  quantitative
agreement with simulation results for times up to the clustering
transition (density instability), for all gas and
fluid densities, and for  small and moderate inelasticities with restitution
coefficient $\alpha > 0.7$.

However, the authors of Ref. \cite{brenig} arrive at quite different conclusions.
From their input data for solving the two-dimensional Boltzmann equation by
DSMC, I extract the following values, relevant for our discussion:
number of particles, $N= 2 \times 10^4$; system size $L = 100 \sigma$
($\sigma$ is a disk diameter) and mean free path $l_0 \simeq 9\sigma$. This
yields the unphysical value for the coverage or area fraction
$\phi = \frac{1}{4} \pi N\sigma^2 / L^2 \simeq 1.57$, which is about twice (!) the
closed packed density of hard disks. According to the Enskog theory 
the value $l_0 \simeq 9\sigma$ corresponds to a coverage of
$\phi \simeq 0.03$, which is not consistent with the authors' data. Moreover
Figure 1 and the text above Eq. (3) of Ref. \cite{brenig} inform the reader 
that
the mean free time is $t_0 \simeq 350 dt$,
where $dt$ is the integration time step in DSMC, and that Haff's law
breaks down at time $ 300 dt \simeq (3/7) t_o$.

We  conclude that the orders of magnitude of the simulation results in
Ref. \cite{brenig} have little relevance for DSMC and  molecular dynamic 
simulations at
realistic values of the system parameter.

\vspace{1cm}

Concerning point (ii): their criticism on granular kinetic theory for three-dimensional
IHS  fluid is based on the following incorrect relation  
for the velocity moments,   $\partial_t \mu_k (t) = - 2\pi 
n \sigma^2 (1-\alpha^k) \mu_{k+1}(t)$, claimed to be valid in the
translationally invariant homogeneous cooling state (HCS). The moments are
defined as,

\begin{equation}
\mu_k(t) \equiv \langle | g_{12} |^k \rangle = \frac{1}{N(N-1)} \int
dx_1 dx_2 |g_{12}|^k f_{0}^{(2)} (x_1 x_2t).
\end{equation}

\noindent
Here $f_0^{(s)} (x_1 \ldots x_s,t)$ are {\it s}-particle distribution functions
which represent the probability density for finding an arbitrary
set of {\it s} particles in the volume element $dx_1 \ldots dx_s$, where
$x_1 = \left\{ \vec{r}_i,\vec{v}_i \right\}$  ($i=1,2,\ldots ,s$) in the HCS. 
These pdf's only depend on relative distances.
Furthermore $g_{ij} \equiv \vec{v}_{ij}\cdot\hat{r}_{ij}$ with
$\vec{v}_{ij} = \vec{v}_i - \vec{v}_j$, and $\hat{r}_{ij}$ is a unit vector
along the lines of centres at  arbitrary nonoverlapping 
positions inside the volume $V = L^3$ of the system. The moments $\mu_k(t)$, 
are intensive quantities, which 
approach for large $V$ the values $(1/4\pi n^2) \times \int d\vec{v}_1
d\vec{v}_2 \int  d\vec{r}_{12} |g_{12}|^k f_0 (\vec{v}_1) f_0 (\vec{v}_2)$,
which are independent of the system size. This holds for
arbitrary long range spatial correlations in $\left[ f_{0}^{(2)} (12) -
f_0 (1) f_0 (2)\right]$, which decay faster than $1/r^a$ with $a > 0$.
The above argument is made to establish that $\mu_k(t)$ is independent of 
the system size.  

To calculate the rate of change $\partial_t \langle A(12) \rangle$ of an
average  pair variable $\langle A(12)\rangle$, as in Eq. (1), I use 
the second equation of the
BBGKY-hierarchy for the reduced distribution functions in terms 
of binary collision operators $\overline{T}_{ij}$
as derived in Ref. \cite{T-op,T-brey,T-twan}, multiply it with $\langle A(12)\rangle /N^2$, 
integrate over $x_1$
and $x_2$, and express it in the adjoint binary collision operator $T_{ij}$,
which reads

\begin{equation}
T_{12} = \sigma^2 \int_{\vec{v}_{12}\cdot\hat{\sigma}_{12}<0}
 d\hat{\sigma}_{12} |\vec{v}_{12}\cdot
\hat{\sigma}_{12} | \delta (\vec{r}_{12} - \sigma\hat{\sigma}_{12})
(b_{\sigma_{12}} -1 ) .
\end{equation}
Here $\hat{\sigma}_{12} = \hat{r}_{12}$ for two spheres in contact, and
$b_{\sigma_{12}} A(\vec{r}_1 \vec{v}_1 \vec{r}_2 \vec{v}_2) = A(\vec{r}_1
\vec{v}_1^{\;\star} \vec{r}_2 \vec{v}_2^{\;\star})$ with
\begin{equation}
\vec{v}_i^{\;\star} = \vec{v}_i - \textstyle{\frac{1}{2}} 
(1+\alpha )(\vec{v}_{ij} \cdot
\hat{\sigma}_{ij})\hat{\sigma}_{ij} .
\end{equation}
In fact, the explicit form of the operator $T_{12}$ is not needed in the 
subsequent arguments, but only that the operator is non-vanishing when 
 particles 1 and 2 are close together.

The result for the rate of change $\partial_t \langle A(12) \rangle$ is then:
\begin{eqnarray}
& &\partial_t \langle A_{12}\rangle = \frac{1}{N^2} \int dx_1 dx_2 f_0^{(2)}
(x_1 x_2) \left( \vec{v}_{12} \cdot \frac{\partial}{\partial \vec{r}_{12}} + 
T_{12}\right) A_{12} + \nonumber \\
& + & \frac{2}{N^2} \int dx_1 dx_2 dx_3 f_0^{(3)} (x_1 x_2 x_3 ) T_{13} A_{12}
\nonumber \\
& \equiv & {\cal K}_1 + {\cal K}_{12} + {\cal K}_{123} .
\end{eqnarray}
Here ${\cal K}_1$ and ${\cal K}_{12}$ represent  the terms containing
respectively $\vec{v}_{12}\cdot\partial /\partial\vec{r}_{12}$ and $T_{12}$
on the first line of (4), and ${\cal K}_{123}$ the term on the second line.
Then, taking $A(12) = |g_{12}|$, ${\cal K}_1$ vanishes, as shown in App.A 
of Ref. \cite{brenig}. 
The $\{ x_1, x_2 \}$-integration in ${\cal K}_{12}$ yields a factor of 
${\cal O}(V)$, as $\delta (\vec{r}_{12} - \sigma\sigma_{12})$
in (2) is constraint to the value $|\vec{r}_{12}| = \sigma$. Consequently, 
the contribution  ${\cal K}_{12}$, introduced in (A7) of Ref. \cite{brenig}. 
is of 
${\cal O}(1/V)$, and vanishes for large systems, and so do the RHS
of Eqs. (A5) and (31) in Ref. \cite{brenig}. 

This does not mean that $\partial_t \mu_k(t)$ in Eq. (31) of 
Ref. \cite{brenig} should be vanishing.
Inspection of ${\cal K}_{123}$ in (4) shows first that this term is
${\cal O}(1)$ for large $V$, because the $\{ \vec{r}_1,\vec{r}_2\}$-integrations
yield a factor $V^2$, because $T_{13}$ poses the constraint $|\vec{r}_{13}| =
\sigma$. For an explicit calculation I insert (2) in (4), and take the
thermodynamic limit. This shows that the contribution of 
$\left[ f_0^{(3)}(123) - f_0(2) f_0^{(2)}(13)\right]$ to ${\cal K}_{123}$ is
of ${\cal O}(1/V)$. With the help of the collision dynamics (1) and (2) I
arrive at the explicit result:
\begin{eqnarray}
{\cal K}_{123} & = & \frac{2\sigma^2}{4\pi n^2}
\int d\vec{v}_1 d\vec{v}_2 d\vec{v}_3
\int^{\prime}_{\vec{v}_{13} \cdot \hat{r}_{13} < 0 }
d\hat{r}_{12} d\hat{r}_{13} f_0 (v_2) f_0^{(2)}
(\sigma\hat{\sigma}_{13}, \vec{v}_1 \vec{v}_3 )
\nonumber \\
& \times & | \vec{v}_{13} \cdot \vec{r}_{13} |
\left\{ | \vec{v}_{12} \cdot \hat{r}_{12} -
\textstyle{\frac{1+\alpha}{2}} (\vec{v}_{13} \cdot \hat{r}_{13})
\hat{r}_{13} \cdot \hat{r}_{12} |^k -
|\vec{v}_{12} \cdot \hat{r}_{12} |^k \right\} .
\end{eqnarray}
The integrations over $\{ \vec{v}_1 \vec{v}_3 
\hat{r}_{13} \}$ are restricted to the precollision  hemisphere, 
$\vec{v}_{13}\cdot \hat{r}_{13} < 0$. Inspection shows that ${\cal K}_{123}$ is
${\cal O}(1)$ and nonvanishing, contrary to what is argued in Appendix A of
Ref. \cite{brenig}.

In summary, the relation $\partial_t \mu_k (t) = - \Gamma_k
\mu_{k+1}(t)$, for the velocity moments claimed to be exact, 
does not hold as it is based
on an incorrect starting point, Eq. (31) of Ref. \cite{brenig}. In deriving 
their Eq.(31) the authors retain a term ${\cal K}_{12}$
which  is of ${\cal O}(1/V)$. This term should be neglected in comparison to
$\partial_t \mu_k(t)$, which is ${\cal O}(1)$ for large $V$.
Therefore, the main simulation and analytic results of this paper
are incorrect.

\end{document}